\documentclass[a4paper]{jpconf}
\usepackage{cite}
\usepackage{amsmath,amsfonts,amsthm,bm, amssymb}
\usepackage{graphicx}
\usepackage{subfig}
\usepackage{lipsum}

\def\epsk{\varepsilon_K}
\def\epsp{\varepsilon^\prime/\varepsilon}
\def\kpnn{K^+\to \pi^+\nu\bar\nu}
\def\kLpnn{K_L\to \pi^0\nu\bar\nu}
\def\kLmm{K_L\to \mu^+\mu^-}
\def\kSmm{K_S\to \mu^+\mu^-}

\def\rtt{\rho_{tt}}
\def\rtc{\rho_{tc}}
\def\rct{\rho_{ct}}

\begin{document}
\title{Kaon processes in general 2HDM}

\author{Girish Kumar}

\address{Department of Physics, National Taiwan University, Taipei 10617,
Taiwan}

\ead{girishk@hep1.phys.ntu.edu.tw}

\begin{abstract}
We discuss new physics (NP) contributions to kaon mixing parameter 
$\varepsilon_K$, direct CP violation parameter
$\varepsilon^\prime/\varepsilon$  of $K\to \pi\pi$,
and rare decays $K^+\to \pi^+\nu\bar\nu$, $K_L\to \pi^0\nu\bar\nu$
and $K_{L, S}\to \mu^+ \mu^- $ in the context of
general two Higgs doublet model. We focus on contributions of
top quark related exotic couplings, and show that simultaneous presence of
flavor conserving and flavor violating interactions can lead to large NP
effects in kaon sector, while being consistent with the stringent constraints
from B physics observables such as $B_{s(d)}$-$\bar B_{s(d)}$ mixing, 
$B_s\to \mu^+ \mu^-$, and $b\to s\gamma$.
We stress on the importance of correlations between
$\varepsilon_K$, $K^+\to \pi^+\nu\bar\nu$ and $B_s\to \mu^+ \mu^-$
that can be exploited to distinguish
the parameter space corresponding to a light (sub-TeV) or heavy (TeV) scale
charged Higgs boson.
\end{abstract}
\section{Introduction}
One Higgs boson particle ($h$), consistent with the prediction of
the Standard Model (SM), has been already found
in 2012 \cite{ATLAS:2012yve,CMS:2012qbp,CMS:2013btf}.
Addition of a second Higgs doublet to the SM  yields a straightforward,
and perhaps the simplest,
extension of the SM known as the two Higgs doublet model (2HDM)~\cite{Lee:1973iz},
where four more physical scalars are predicted:
a CP-even neutral scalar ($H$), a CP-odd neutral scalar ($A$),
and two charged scalars ($H^\pm$).
In this article, we consider a general 2HDM (of Type-III~\cite{Hou:1991un})
which contains flavor changing neutral couplings (FCNC)
at the Lagrangian level itself (see Ref.~\cite{Branco:2011iw} for a comprehensive review).
In the so-called Higgs basis \cite{Georgi:1978ri,Lavoura:1994fv,Botella:1994cs}
where only one of the scalar doublet develops vacuum expectation value,
the interactions between physical scalars and SM fermions are described
by the following Lagrangian \cite{Davidson:2005cw,Hou:2017hiw},
\begin{align}
\mathcal{L}_Y = 
 -  \frac{1}{\sqrt{2}} \sum_{f = u, d, \ell} \bar f_{i}
    \Big[\big(\lambda^f_i \delta_{ij} s_\gamma 
            + \rho^f_{ij} c_\gamma\big) h 
      + \big(\lambda^f_i \delta_{ij} c_\gamma
            - \rho^f_{ij} s_\gamma\big) H
      - i\,{\rm sgn}(Q_f) \rho^f_{ij} A\Big] P_R\, f_{j} \nonumber\\
 - \bar{u}_i\left[(V\rho^d)_{ij}P_R-(\rho^{u\dagger}V)_{ij}P_L\right]d_j H^+
 - \bar{\nu}_i\rho^\ell_{ij} P_R \, \ell_j H^+
 +{h.c.},
\label{eq: Lag}
\end{align}
where $i, j$ are generation indices, $\lambda_i\,(= 
\sqrt{2} m_i/v)$ are the Yukawa coupling in the SM, and 
$\rho^f_{ij}$ denote NP couplings; 
$V$ is the Cabibbo-Kobayashi-Maskawa (CKM)
matrix and $P_{R/L} = (1 \pm \gamma_5)/2$ are chirality 
projecting operators.
The shorthand notation $c_\gamma$ ($s_\gamma$) denotes the cosine (sine) of
the mixing angle $\gamma$ between  $h$ and $H$.
It is worth mentioning that  the \emph{alignment limit} (i.e. $ c_\gamma \to 0$) helps
in suppressing the flavor violating decays of $h$ boson, $h\to f_i f_j$ ($i\ne j$).

The NP couplings $\rho_{ij}$ are generic in size and need to be constrained from
experimental data.
One of the strongest constraints on these couplings comes from
the precision measurements of B physics observables;  data on neutral $B$ meson mixing
and rare decays such as $B_s\to \mu^+\mu^-$ and $B\to X_s\gamma$ provides severe constraints
on  the parameter space of 2HDM (for example, see \cite{Crivellin:2013wna}).
In this work, we discuss the important role of the kaon sector
in the probe of general 2HDM, and show that in many cases kaon processes
can provide far better constraints.

To keep our numerical analysis concise, we focus on certain
top-related NP Yukawa couplings only, and assume matrices $\rho^{d, \ell}$
to be vanishing. More explicitly, we use the following ansatz
\begin{align}\label{eq: rhos}
	\rho^u \equiv \begin{pmatrix}
		0 & 0 & 0\\0 & 0 & \rct \\0 & \rtc & \rtt\\
	\end{pmatrix}, 
	\quad \rho^{d} =\rho^\ell = 0,
\end{align}
while for masses of exotic scalars,
we consider $400$ GeV and $1000$ GeV as two reference values.

\begin{figure}[h]
\begin{center}
\subfloat[]{\includegraphics[width=0.2\textwidth, height=0.092\textheight]{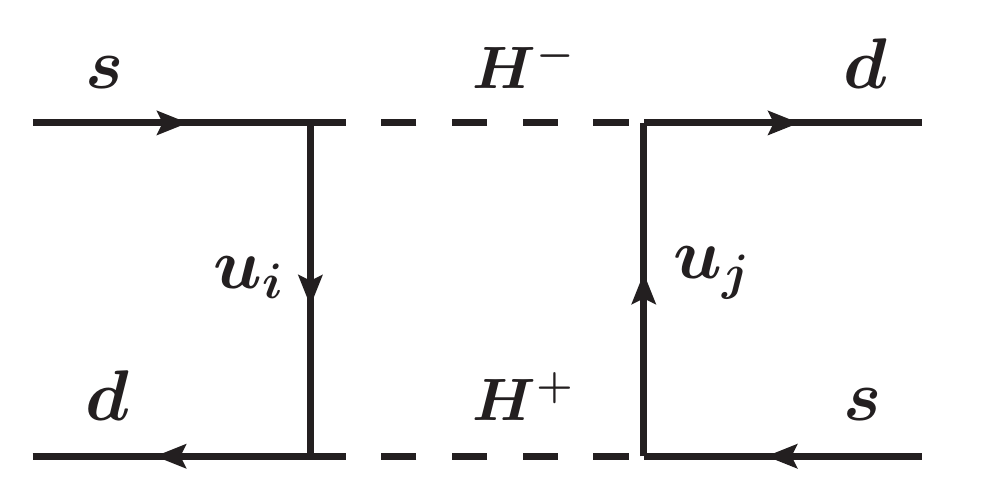}
\includegraphics[width=0.2\textwidth, height=0.092\textheight]{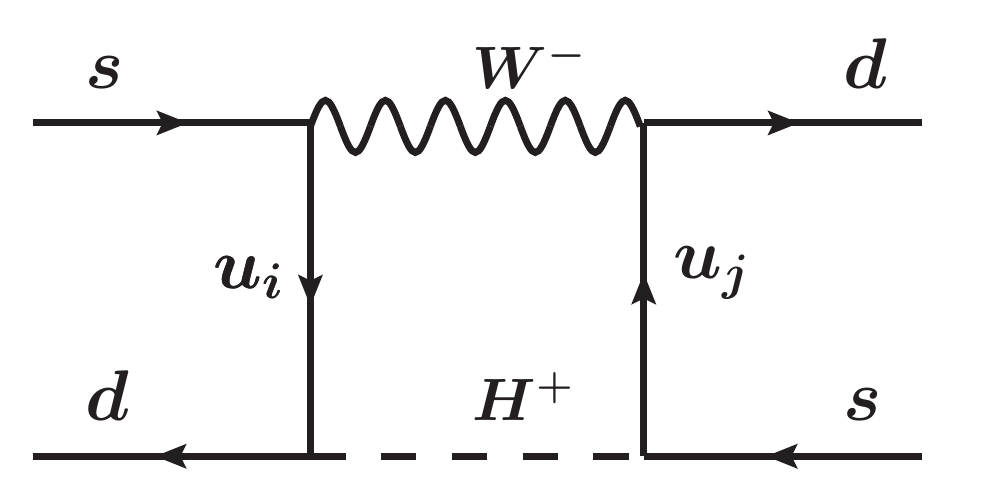}\label{fig: Feyn_a}}
\quad \quad 
\subfloat[]{\includegraphics[width=0.2\textwidth]{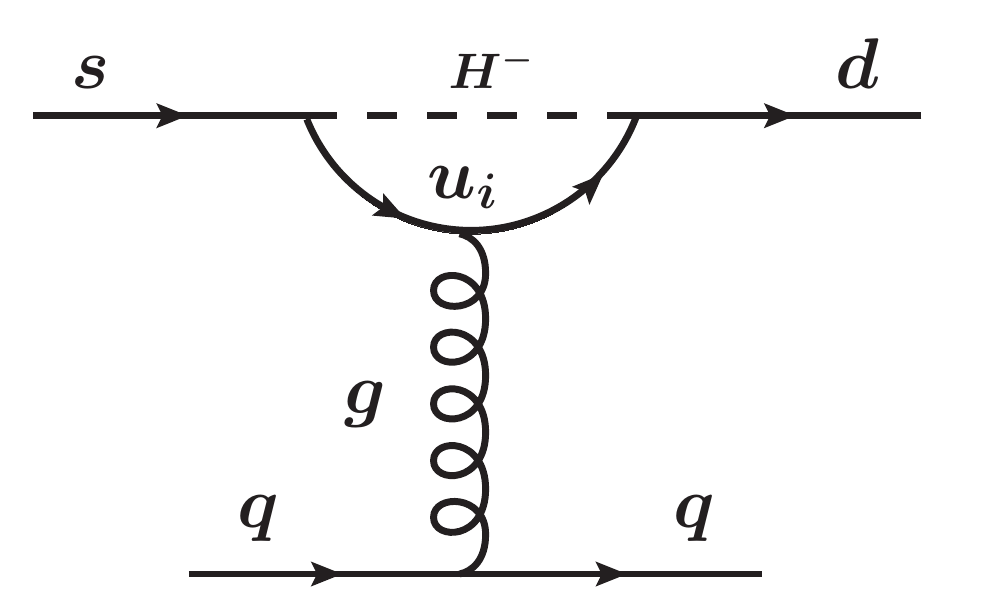}\label{fig: Feyn_b}}
\quad \quad 
\subfloat[]{\includegraphics[width=0.2\textwidth]{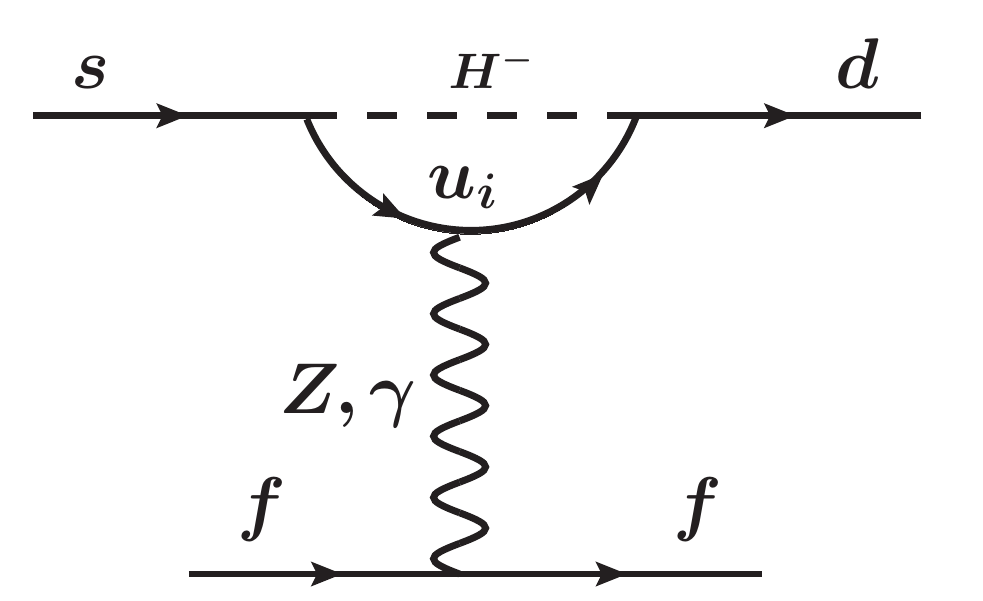}\label{fig: Feyn_c}}
\end{center}
\caption{Representative Feynman diagrams for $K^0$-$\bar K^0$  mixing (a);
$s \to d q\bar q$ (b and c) ; and $s \to d \nu\bar\nu$ and  $s \to d \ell\bar\ell$  (c), where for  $f= \nu$, only $Z$-penguin contributes. }
\label{fig: Feyn}
\end{figure}

In our analysis, we consider following kaon observables:
(i)  $K^0$-$\bar K^0$  mixing parameter $\epsk$,
(ii) direct CPV parameter $\epsp$ of $K\to\pi\pi$,
(iii) rare semileptonic decays $\kpnn$ and $\kLpnn$,
(iv) rare decays $\kLmm$ and $\kSmm$.
With NP couplings as defined in Eq.~\eqref{eq: rhos},
the NP contributions to above  processes arise from $H^+$ mediated loop diagrams
as shown in Fig.~\ref{fig: Feyn}. The corresponding theoretical expressions can be
found in our paper~\cite{Hou:2022qvx}.

%
\section{Results}
\subsection{Constraints from B physics and $\epsk$}
\begin{figure}[b]
\begin{center}
\includegraphics[width=5.cm,height=4.cm]{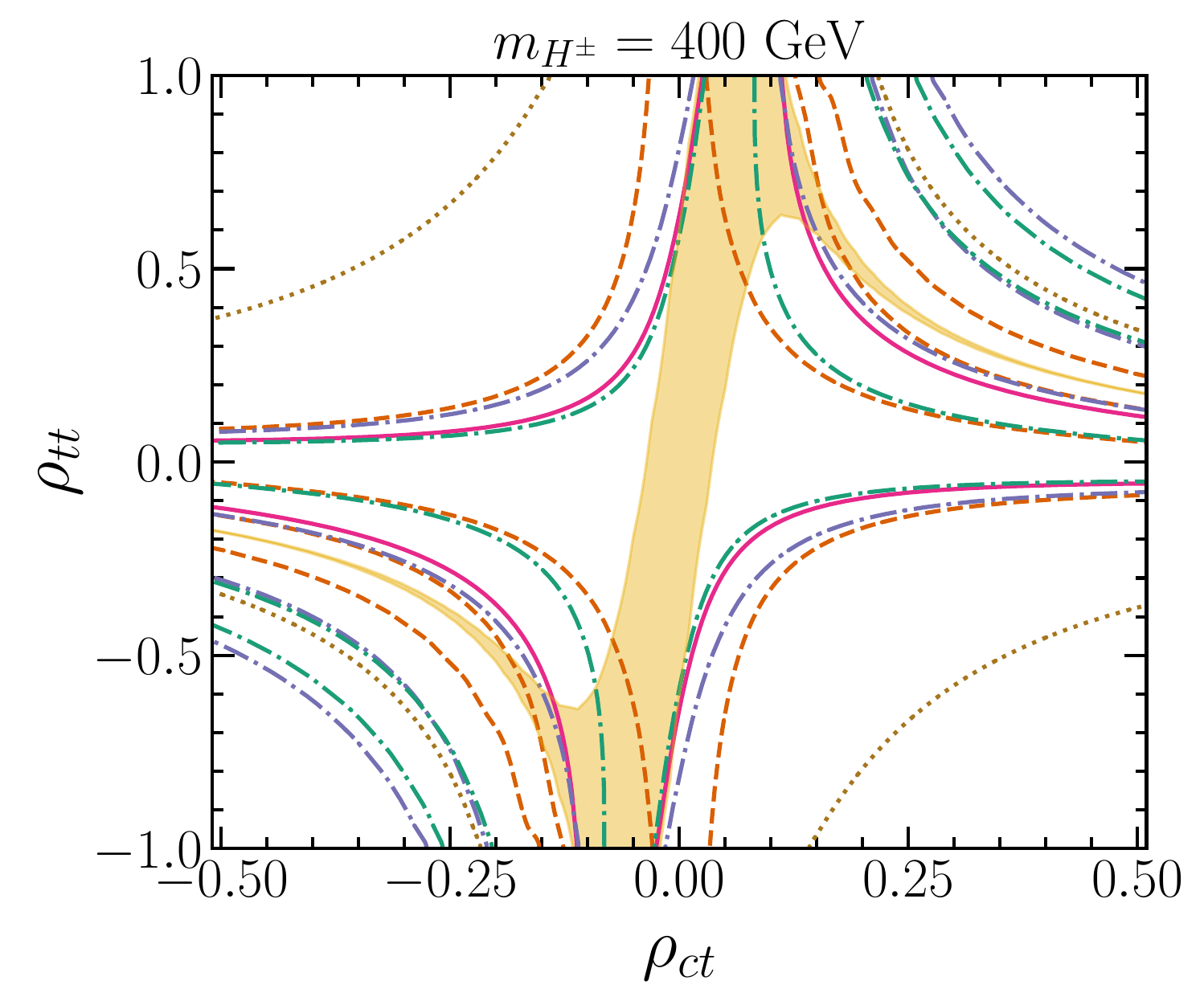}
~~\includegraphics[width=6.7cm,height=4.cm]{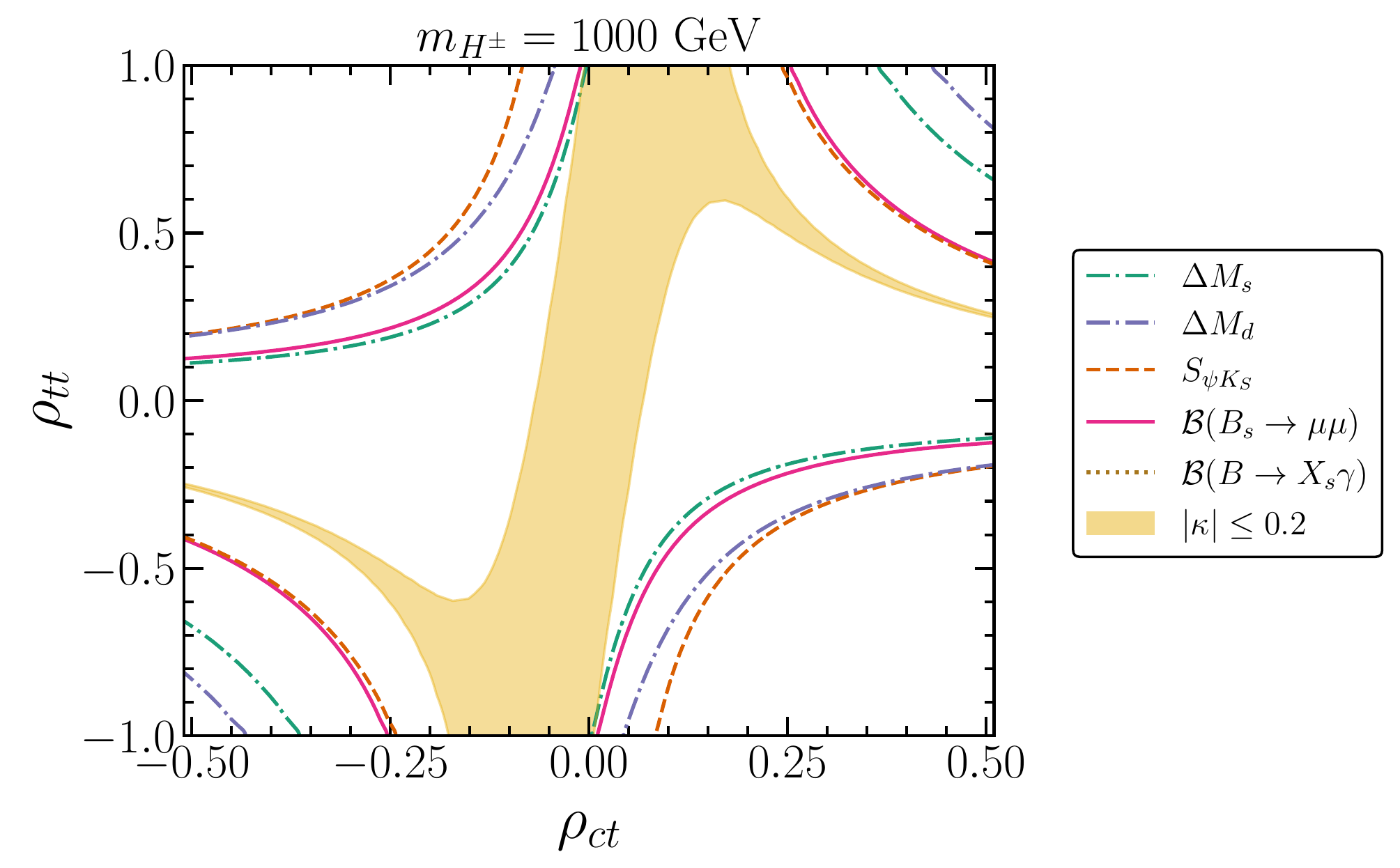}
\end{center}
\caption{Flavor constraints from B physics and $\epsk$.}
\label{fig: B-epsK}
\end{figure}
We first discuss constraints on $\rho_{ij}$ couplings from B physics. In Fig.~\ref{fig: B-epsK},
we show constraints from neutral $B_q$ ($q=s, d$)  mass difference $\Delta M_{q}$,
 mixing-induced CP-asymmetry $S_{\psi K_S}$ in $B_d\to \psi K_S$, and 
branching ratios of $B_s\to \mu^+\mu^-$ and $B\to X_s\gamma$ in the plane
of couplings $\rho_{tt}$ and $\rho_{ct}$ for $m_{H^+}= 400$ and $1000$ GeV.
We note that B physics observables give strong bounds on $\rho_{tt}$ and $\rho_{ct}$,
but still allow them to be  large, especially when one of the
coupling is vanishing.
In passing, we mention that constraints on the coupling $\rho_{tc}$
are relatively very weak as this coupling is associated with the (small)
$H^+$-charm quark loop. 

Next we include constraints from $\epsk$.
Defining $\epsk^{\rm NP} \equiv \kappa \times 10^{-3}$ as NP contribution to $\epsk$,
the current data allows for $-0.2 \le \kappa \le 0.2$ ~\cite{Aebischer:2020mkv}.
This constraint is shown as yellow region in 
Fig.~\ref{fig: B-epsK}, which highlights the remarkable ability of $\epsk$,
compared to B physics observables,
in constraining  $\rho_{ct}$. However, it is worth noting 
that for heavy $H^+$ case, the B physics constraints become weak and $\epsk$
admits sizable $\rho_{ct}$, as can be seen from Fig.~\ref{fig: B-epsK} (right). 

\subsection{$\kpnn$ as sensitive probe of a heavy $H^+$}
We now discuss results for rare decays $\kpnn$ and $\kLpnn$.
The former has been measured by the NA62 experiment \cite{NA62:2021zjw}
while for the latter the KOTO experiment has set a $90\%$ C.L.
bound~\cite{KOTO:2020prk}. These corresponding results are:
  ${\cal B}(\kpnn)_{\rm NA62} = (10.6^{+4.0}_{-3.4} \pm 0.9)\times 10^{-11}$, ${\cal B}(\kLpnn)_{\rm KOTO} < 4.9 \times 10^{-9}$.
For the corresponding  values in the SM, we find
	${\cal B}(\kpnn)_{\rm SM} = (9.07 \pm 0.82)\times 10^{-11}$, ${\cal B}(\kLpnn)_{\rm SM} = (3.24 \pm 0.36) \times 10^{-11}$.
\begin{figure}[b]
\centering
\includegraphics[width=5.1cm, height=4cm]{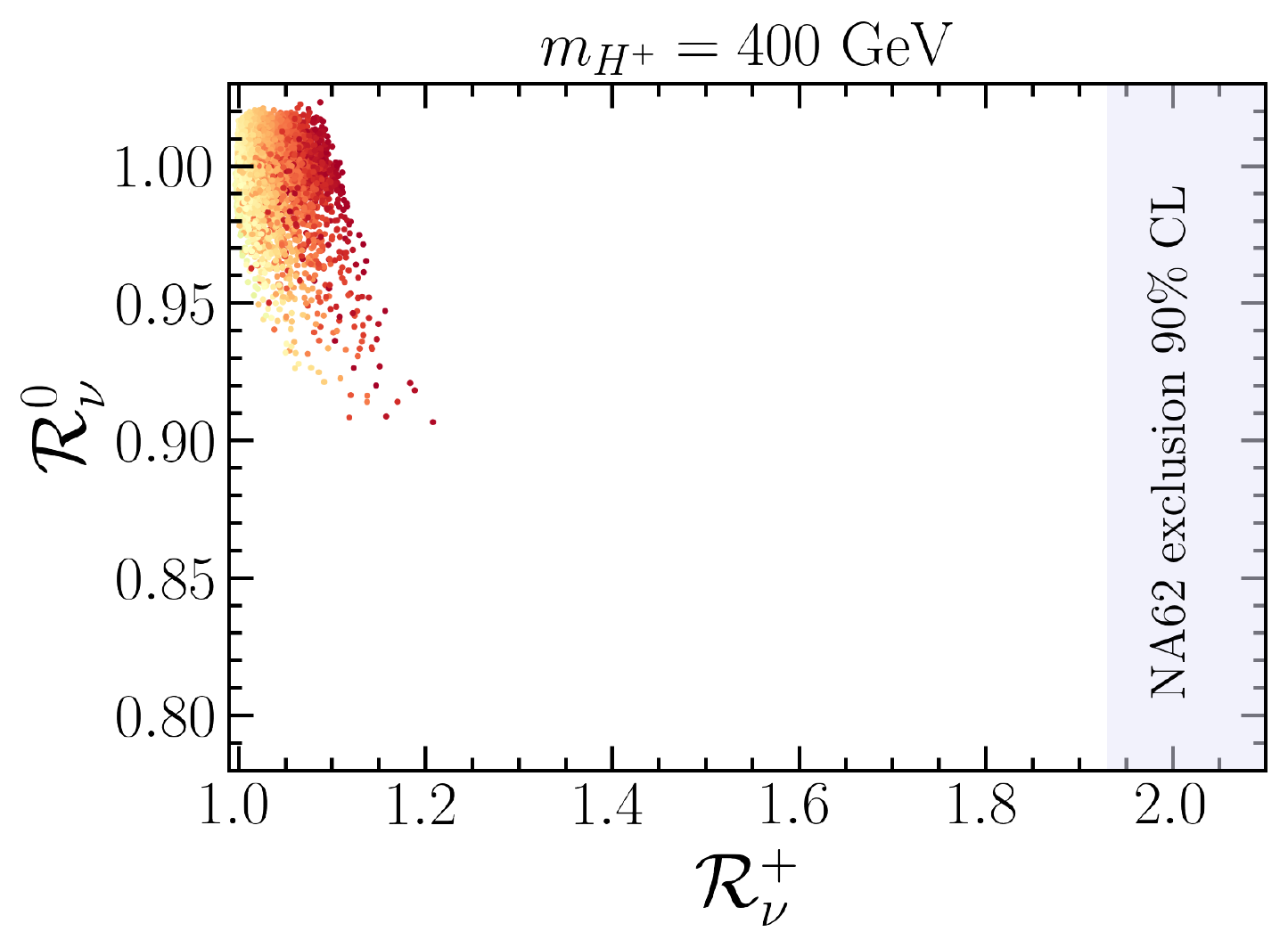}~~~\includegraphics[width=6.8cm, height=4cm]{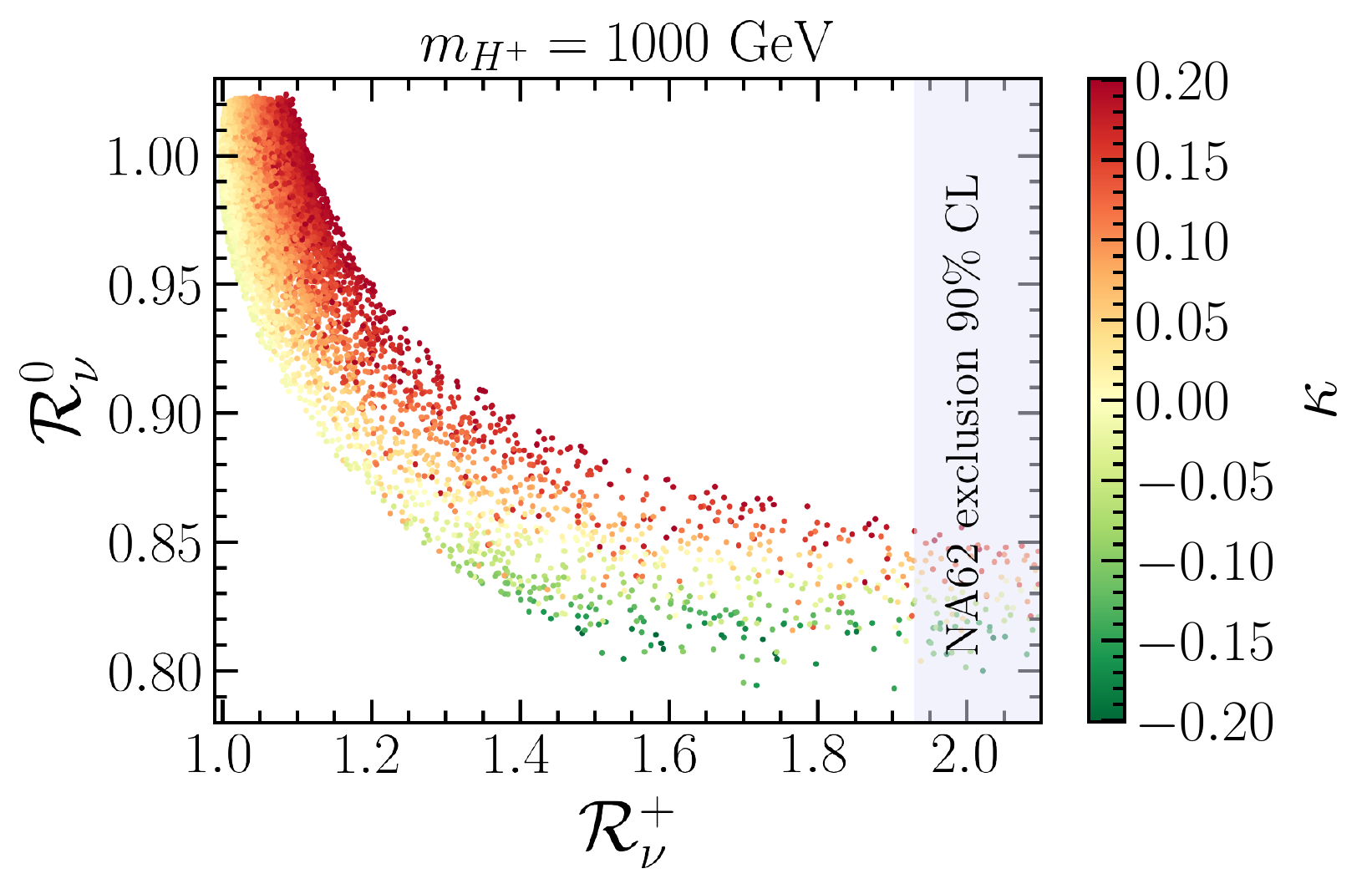}
\caption{\label{fig: Rnu} Correlation between $\kpnn$, $\kLpnn$, and 
$\epsk$.}
\end{figure}

To discuss results for  $\kpnn$ and $\kLpnn$, we introduce following ratio,
\begin{align}
  {\cal R}_{\nu}^{+}
  = \frac{{\cal B}(K^+ \to \pi^+ \nu\bar \nu)}
       {{\cal B}(K^+ \to \pi^+ \nu\bar \nu)_{\rm SM}},\quad 
  {\cal R}_{\nu}^{0}
  = \frac{{\cal B}(K_L \to \pi^0 \nu\bar \nu)}
       {{\cal B}(K_L \to \pi^0 \nu\bar \nu)_{\rm SM}},
\end{align}
which in the SM, by definition, are unity. 
In Fig.~\ref{fig: Rnu}, we show results for ${\cal R}_{\nu}^{+}$ and ${\cal R}_{\nu}^{0}$.
To obtain these results, we define
$\rho_{ij} \equiv |\rho_{ij}| \exp(i \phi_{ij})$,
and scan over
$|\rtt|,\,|\rtc| \in [0, 1]$, $\phi_{tt},\,\phi_{tc} \in [-\pi, \pi ]$;
 $|\rct| \in [0, 0.3],\, \phi_{ct} \in [-\pi, \pi ]$.
The allowed points are obtained after imposing B physics and $\epsk$ constraints.
We note that for light $H^+$ case (left),  
${\cal R}_{\nu}^{+}$ gets enhanced up to $\sim 20\%$
while ratio  ${\cal R}_{\nu}^{0}$ is suppressed up to $\sim 10\%$ compared to the SM value.
For the heavy $H^+$ case (right),
the NP effects are relatively larger: ${\cal R}_{\nu}^{+}$ can easily saturate the NA62 limit,
while ${\cal R}_{\nu}^{0}$ can be up to $\sim 20\%$ suppressed.
At first glance, the larger NP effects in $K\to \pi\nu\bar\nu$
for heavy $H^+$ may appear surprising, but this can be understood from the following.
In general 2HDM, these decays receive dominant NP contribution from  $Z$-penguin diagram
with $H^+$-top  loop (shown in Fig.~\ref{fig: Feyn_c}).
The corresponding NP Wilson coefficient, normalized by the SM one,
is proportional to the following combination~\cite{Hou:2022qvx},
\begin{align}\label{eq: s2dnunu}
 \left(\rtt   + \frac{V_{cs}^*}{V_{ts}^*}\rct\right)
   \left(\rtt^* + \frac{V_{cd}}{V_{td}}\rct^*\right)
  G_Z({m_t^2/m_{H^+}^2)},
\end{align}
where $G_Z(x)$ is the loop function given in Ref.~\cite{Hou:2022qvx}.
From Eq.~\eqref{eq: s2dnunu}, we note that $\rct$
terms are enhanced by CKM factors, $V_{cs}^\ast/V_{ts}^\ast\simeq -23.5 - 0.46\, i$, and
$V_{cd}/V_{td}\simeq-22.8 - 9.4\, i$, respectively. 
Furthermore, it can be shown that dominating part of total $H^+$ contribution to
$s\to d \nu\bar\nu$ processes is  CP conserving (for details, see ref.~\cite{Hou:2022qvx}).
The CKM-enhanced sensitivity of $\kpnn$ to $\rct$, coupled with the fact that flavor constraints on $\rct$ become weak for heavy $H^+$ (see Fig.~\ref{fig: B-epsK}),
explains why $\kpnn$ is an excellent probe of heavy $H^+$.

\subsection{Correlation of $\kpnn$ with $B_s\to\mu^+\mu^-$}
\begin{figure}[t]
\centering
\includegraphics[width=5.14cm, height=4.cm]{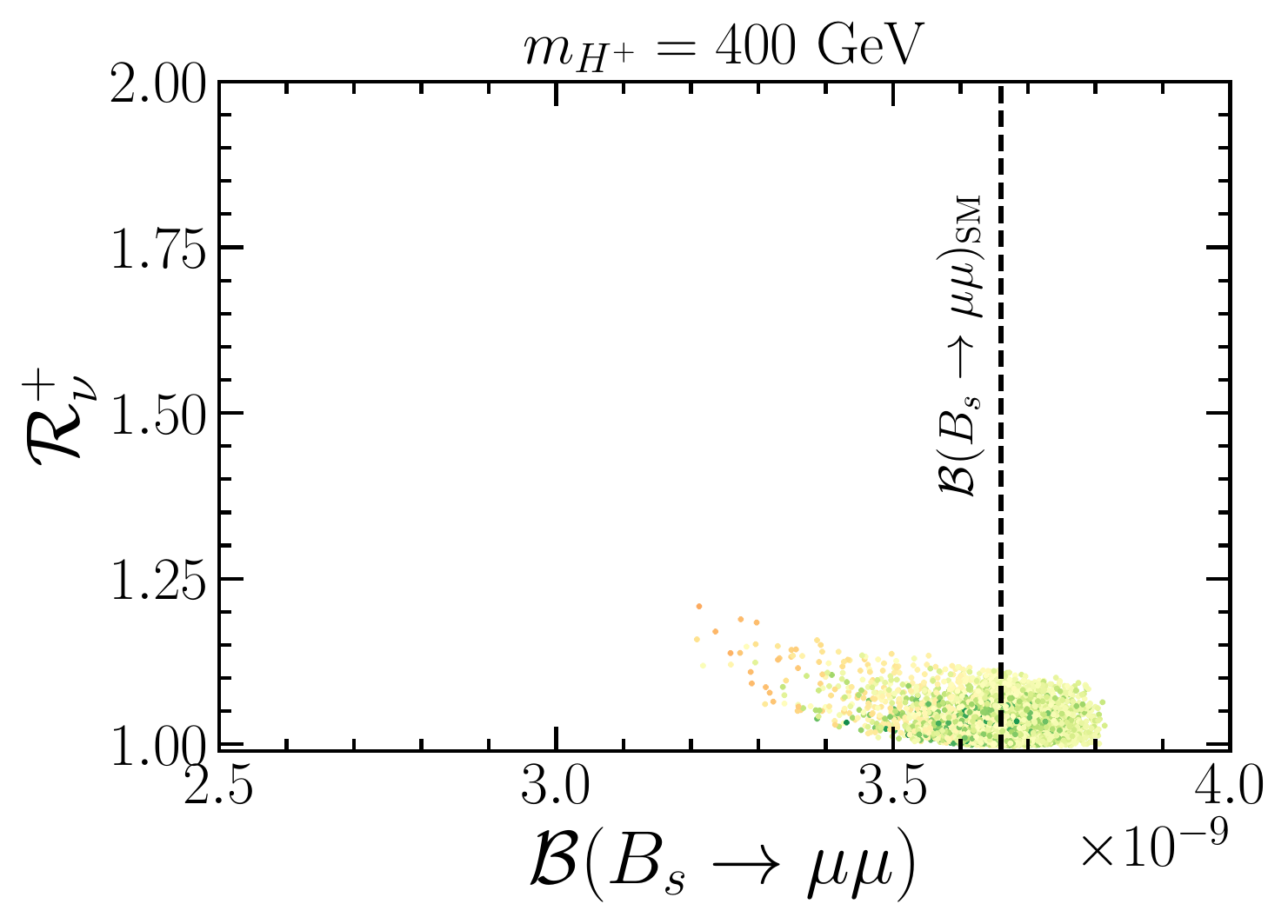}~~~\includegraphics[width=6.3cm, height=4.cm]{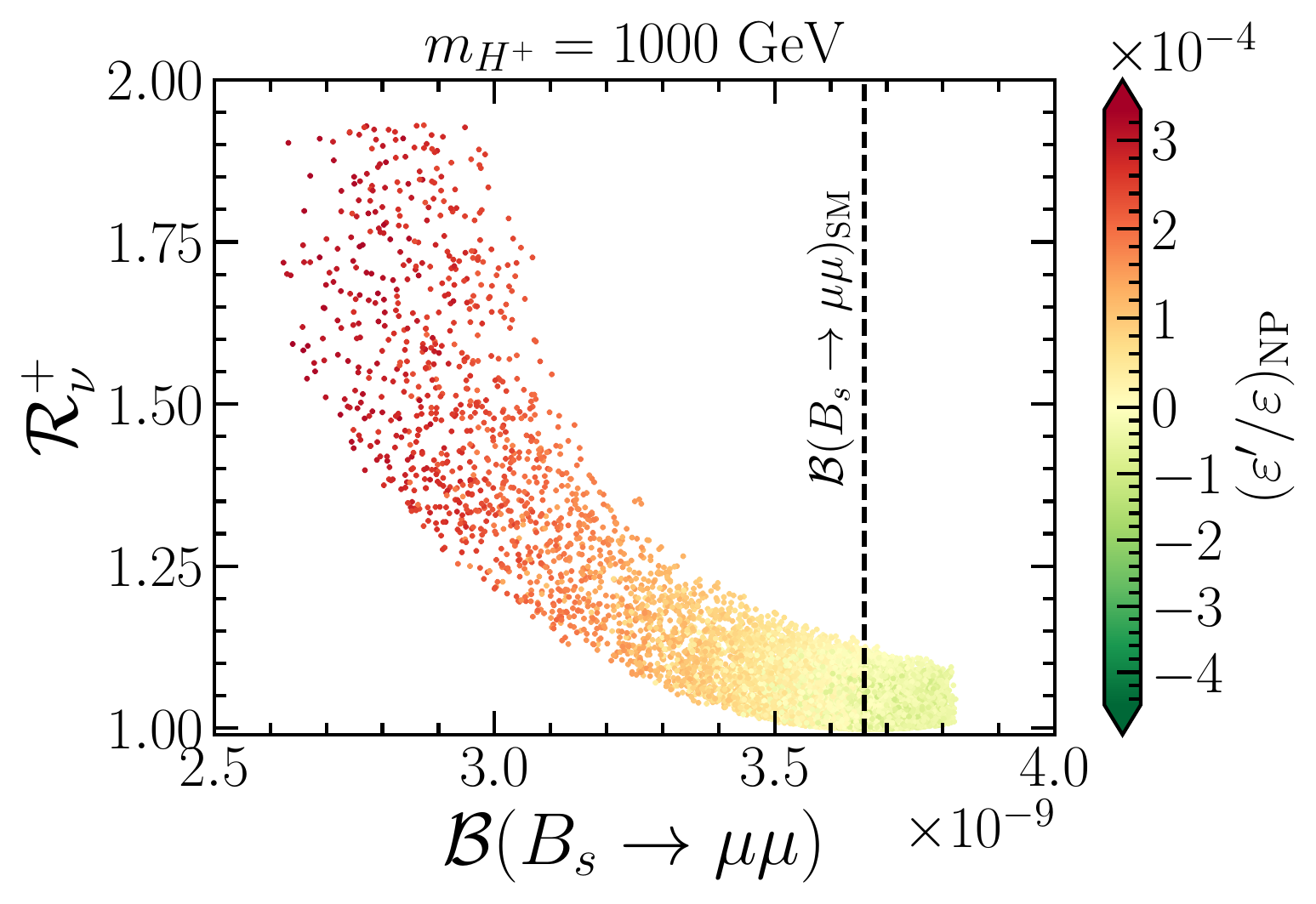}
\caption{\label{fig: Rnu-Bsmm} Correlation between $\kpnn$,  $B_s\to \mu^+\mu^-$, and $(\epsp)_{\rm NP}$. The dashed line indicates the central value of ${\cal B}(B_s \to \mu^+\mu^-)$ in the SM.}
\end{figure}

Let us now discuss the correlation of NP effects 
in ${\cal R}_\nu^{+}$ and ${\cal B}(B_s \to \mu^+\mu^-)$, which provide another crucial
probe of a heavy $H^+$.
In Fig.~\ref{fig: Rnu-Bsmm}, we show the scatter of allowed points in the plane
of $B_s\to\mu^+\mu^-$ and $\kpnn$ as function of $(\epsp)_{\rm NP}$, where we note
that enhancement of ${\cal R}_\nu^{+}$ is correlated with the suppression of
${\cal B}(B_s \to \mu^+\mu^-)$. For light $H^+$ case (left),
the combined constraints from B physics and $\epsk$
restrict $B_s\to\mu^+\mu^-$ within $2\sigma$ range of SM value, $(3.66 
\pm 0.14) \times 10^{-9}$~\cite{Beneke:2019slt}, while ${\cal R}_\nu^{+}$ 
can be up to $\sim 20\%$ enhanced. But in the heavy $H^+$ case (right),
the anti-correlation of $\kpnn$ with $B_s\to\mu^+\mu^-$ is clearly noticeable:
the higher values of ${\cal R}_\nu^{+}$ are correlated with more suppressed values
of ${\cal B}(B_s \to \mu^+\mu^-)$.
As indicated in Fig.~\ref{fig: Rnu-Bsmm}, contribution to $(\epsp)_{\rm NP}$
lie in the range of $-4\times 10^{-4}$ to $3\times 10^{-4}$,
 where larger values of $(\epsp)_{\rm NP}$ are correlated with enhanced
${\cal R}_\nu^{+}$ and suppressed
$B_s\to\mu^+\mu^-$, and vice versa. Note that  above values of $(\epsp)_{\rm NP}$
are compatible with the current data \cite{Aebischer:2020jto}.

Before concluding, we comment on rare decays $K_{L, S}\to \mu^+\mu^-$.
We find that large uncertainties 
associated with theoretical determination of ${\cal B}(K_{L}\to \mu^+\mu^-)$
makes it a less effective probe
of $H^+$ in comparison  to $\kpnn$, while for $K_S\to \mu^+\mu^-$,
we find  NP effects to be less than $2\%$~\cite{Hou:2022qvx}.

\section{Conclusions}
In this work, we investigated contributions of top-related NP Yukawa couplings
of general 2HDM to $K^0$-$\bar K^0$ mixing,
$\epsp$, and rare decays $\kpnn$, $\kLpnn$, and $K_{L, S}\to \mu\mu$.
We found that $\epsk$, in comparison to B physics observables, provides better constraint
on the off-diagonal coupling $\rct$. For sub-TeV values of $m_{H^+}$,
we found that current B physics
and $\epsk$ data allow only mild NP effects in rare kaon decays.
However, if $H^+$ is  TeV-scale heavy then  flavor constraints become weak,
and, thanks to large CKM enhancement
of $\rct$ terms, substantial NP effects in $\kpnn$ are possible.
Another important result is that NP effects
 in $\kpnn$ and $B_s\to\mu^+\mu^-$ are found to be anti-correlated,
 which can be exploited to probe the scale of $H^+$.

\ack
It is a pleasure to thank the organizers of Kaon 2022 conference for
the kind invitation to give  the talk.
I also wish to thank Prof. George W.-S. Hou for collaboration on
Ref.~\cite{Hou:2022qvx}, which this article is based on. This work is supported by
NSTC 111-2639-M-002-002-ASP of Taiwan.

\section*{References}
\bibliography{main}
\end{document}